\pdfoutput=1
\documentclass[preprint,aps,superscriptaddress]{revtex4}
\usepackage{graphicx,amssymb,bm,natbib,amsfonts,amsmath}
\usepackage{hyperref}

\begin{document}

\title{Origin of the energy bandgap in epitaxial graphene}

\author{S.Y. Zhou}
\affiliation{Department of Physics, University of California,
Berkeley, CA 94720, USA}
\affiliation{Materials Sciences Division,
Lawrence Berkeley National Laboratory, Berkeley, CA 94720, USA}

\author{D.A. Siegel}
\affiliation{Department of Physics, University of California,
Berkeley, CA 94720, USA}
\affiliation{Materials Sciences Division,
Lawrence Berkeley National Laboratory, Berkeley, CA 94720, USA}

\author{A.V. Fedorov}
\affiliation{Advanced Light Source, Lawrence Berkeley National Laboratory, Berkeley, California 94720, USA}

\author{F. El Gabaly}
\affiliation{Materials Sciences Division,
Lawrence Berkeley National Laboratory, Berkeley, CA 94720, USA}

\author{A. K. Schmid}
\affiliation{Materials Sciences Division,
Lawrence Berkeley National Laboratory, Berkeley, CA 94720, USA}

\author{A.H. Castro Neto}
\affiliation{Department of Physics, Boston University, 590 Commonwealth Avenue, Boston, Massachusetts 02215, USA}

\author{A. Lanzara}
\affiliation{Department of Physics, University of California,
Berkeley, CA 94720, USA}
\affiliation{Materials Sciences Division,
Lawrence Berkeley National Laboratory, Berkeley, CA 94720, USA}

\date{\today}

\maketitle

We show here that the argument presented by Rotenberg {\it et al.} in the preceding comment to account for the energy gap reported by us \cite{ZhouNatMat} in epitaxial graphene on 6H-SiC is unfounded. In our view, the effects of modulations on the lateral structure of graphene films alone cannot account for the large gap observed in films with single-layer graphene terraces exceeding 150 nm in size.

It is no surprise that a gap can be induced in graphene by confined geometries such as nanoribbons or quantum dots as this has been predicted and observed experimentally \cite{Nakada, Han, Ohta}.  However, the data presented below show that this is not the main mechanism behind the gap opening in epitaxial graphene. 

To investigate the role of electron confinement in opening the gap, in Figure 1 we study the dependence of the gap size on the average size of the graphene terraces and compare it with the results for nanoribbons \cite{Han}. The size of the terraces (controlled by systematically varying the degree of under-annealing) and the size of the excitation gap are directly measured on the same samples by low energy electron microscopy (LEEM) and angle resolved photoemission spectroscopy (ARPES), respectively. The size of the single layer graphene terraces in the LEEM images (panels a-c) is extracted by first identifying outlines of the terraces, as described by Ohta {\it et al.} \cite{Ohta}, and then by quantifying the average size by drawing straight lines in the images (we chose the image diagonal) and measuring the linear widths of the terraces crossed by the lines \cite{Wang}.  The gap size is extracted from the ARPES spectra at the K point (panels d-f) as described by Zhou {\it et al.} \cite{ZhouNatMat}.  The direct comparison between LEEM and ARPES data allows one to readily correlate the size of the excitation gap and the width of the photoemission features to the average size of the single layer graphene terraces.

\begin{figure*}
\includegraphics[width=15.8 cm] {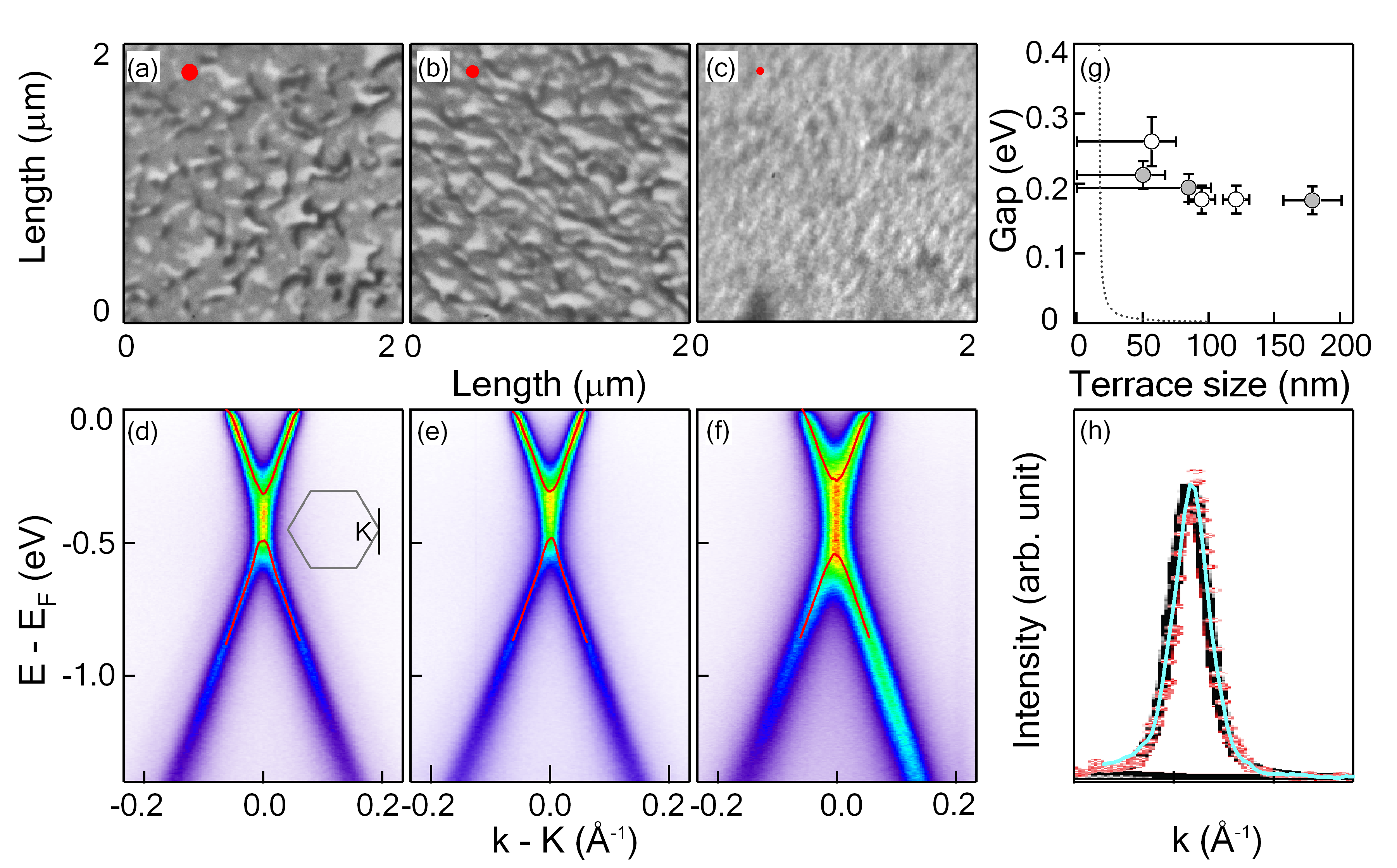}
\label{Figure 1}
\caption{Direct correlation of graphene terrace sizes measured by LEEM and the gaps measured from ARPES shows that quantum confinement is not the main mechanism for the gap opening. (a-c) LEEM images (smooth background subtracted), taken at electron energy of 6.6 eV (a,b) and 7.1 eV (c) show the surface topography graphene films for three characteristic samples prepared under different conditions. White, gray and black colors in panels (a-c) represent regions of buffer layer, single layer and bilayer graphene respectively.  Diameters of red circles in panels (a-c) represent the average size of the single layer graphene terraces.  In panel c, small size of the single layer graphene terraces is not well resolved due to limited image resolution, and our size measurement must be viewed as an upper limit. (d-f) Corresponding ARPES data are taken through the K point (see vertical line in the inset of Fig.~1(d)). Red lines are dispersions extracted by fitting energy distribution curves.  (g) Plot of the gap size extracted from the ARPES data as a function of the single layer graphene terrace size.  The open symbols are extracted from data shown in panels (a-f), and the filled symbols are taken from three additional samples not shown.  The dotted line is the gap size in graphene nanoribbons due to quantum confinement taken from Ref.\onlinecite{Han}. (h) Comparison of our momentum distribution curves at EF from panel b (cyan lines) with data from Rotenberg {\it et al.} (red symbols).}
\end{figure*}

The main result is summarized in panel g. Although as expected, a slight increase in the magnitude of the gap for the smallest terrace size is observed \cite{Nakada, Son, Han}, it is clear that in samples with terrace size larger than 80 nm, the gap does not change within our experimental error bar and is finite at around 180 meV.  In contrast, the predicted and observed quantum confinement gap for exfoliated samples (dashed line) almost vanishes for ribbons larger than 30 nm and, clearly, cannot account for what is observed in our data.  These results strongly indicate that the gap is an intrinsic property of epitaxial graphene.  This conclusion is further supported by Kim {\it et al.} who independently performed rigorous ab-initio calculations and predicted a gap at the Dirac point in epitaxial graphene on SiC \cite{Kim}.  Their calculated spectra are nearly identical to the ARPES data \cite{ZhouNatMat, Bostwick} and show a perfect agreement both for the magnitude of the gap and the finite intensity inside the gap due to the presence of midgap states. 

It is important to note that as part of our analysis described above we examined large domain samples with spectral features very similar to those studied by Bostwick et al. (see comparison in panel h).  The observation of a finite gap in this sample and the similarity with the spectra calculated by Kim {\it et al.} \cite{Kim} casts doubt on the explanation by Bostwick {\it et al} \cite{Bostwick}, which is based on electron-plasmon interaction \cite{Bostwick}.  Detailed discussion of the gap vs. the electron-plasmon scenario is presented in Ref.~\onlinecite{Kim} and in Ref.~\onlinecite{ZhouPhysicaE}.

Finally, we would like to point out that the lack of evidence of sixfold symmetry breaking in scanning tunnelling microscopy (STM) \cite{Brar} does not in our view
contradict the intrinsic nature of the gap \cite{ZhouNatMat, Kim}.  The STM data are taken far away from the Dirac point, where it is known that the eff ect of the symmetry breaking is much
weaker (see also ref.~\onlinecite{ZhouNatMat}, Fig. 4), and difficult to observe. Therefore this result does not contradict the ARPES data and the gap scenario presented by us \cite{ZhouNatMat}.

In summary, by showing data from samples with different graphene terrace sizes, we prove that the excitation gap at the Dirac point in the single layer graphene grown on 6H-SiC is not due to quantum confinement as the preceding comment suggests, but is an intrinsic property of epitaxial graphene.  Regardless of whether the gap is induced by symmetry breaking or not, its observation is an important step in understanding the physics of low-dimensional epitaxial graphene \cite{Novoselov}.

\begin{acknowledgments}
We thank K.F. McCarty for experimental assistance with the LEEM experiments.
\end{acknowledgments}

\end{document}